\def\lam{$\Lambda$}
\def\he5l{$^{5}_\Lambda$He}
\def\li6l{$^{6}_\Lambda$Li}

\def\pnpl{$p n  \rightarrow p \Lambda  $}
\def\c12{$^{12}$C}
\def\c12l{$^{12}_\Lambda$C}

\def\pikli6{$^{6}{\rm Li}(\pi^+,K^+)$}

\documentstyle[preprint,eqsecnum,aps,epsf]{revtex}
\tightenlines

\begin{document}
\draft
\title{ Spin observables in the $pn \rightarrow p \Lambda$ reaction }
\author{ H. Nabetani$^1$, T. Ogaito$^2$, T. Sato$^1$ and T. Kishimoto$^1$ }
\address{1
Department, of Physics, Osaka University, Toyonaka, Osaka, 560-0043, Japan}
\address{2
Department of Physics, Fukui Medical University,
Fukui, 910-1193, Japan }
\date{\today}
\maketitle
\begin{abstract}
	The T matrix of the \pnpl\ reaction, which is a strangeness 
changing weak process, is derived.  The explicit formulas of the spin 
observables are given for s-wave $p \Lambda$ final states which 
kinematically corresponds to inverse reaction of the weak nonmesonic decay 
of \lam\ hypernuclei.   One can study interferences between amplitudes of 
parity- conserving and violating, spin- singlet and triplet and isospin- singlet 
and triplet.   Most of them are not available in the study of the 
nonmesonic decay.   They clarify structure of the reaction and constrain 
strongly theoretical models for weak hyperon nucleon interaction.  
\end{abstract}

\narrowtext

The nonmesonic decay (NM-decay) of \lam\ hypernuclei ($N \Lambda 
\rightarrow N N$) is the only process through which one has studied the 
strangeness changing weak baryon baryon (BB) interaction so far.  Since 
the weak interaction does not conserve parity, a complete understanding of 
the process needs study of both parity- conserving and violating 
amplitudes.  For the weak nucleon nucleon (NN) interaction, which is a 
non-strange part of the weak BB interaction, one can study only its 
parity-violating part, because the parity conserving part is completely 
masked by the strong interaction.   Both amplitudes can be studied with the 
weak NM-decay since no strong interaction can change flavor.   The weak NN 
and hyperon nucleon (YN) interaction can be understood in an unified way 
based on the SU$_F$(3) symmetry.  The study of the NM-decay is thus 
interesting and informative.  One can study kinematically limited region 
of the weak YN interaction using NM-decay.  In the present letter we show 
that many spin observables measurable in the inverse \pnpl\ reaction will 
open a new opportunity to study the weak YN interaction generally.

Until recently experimental data were available on total- and 
partial-decay rates of the NM-decay.  Proton stimulated decay 
$(p \Lambda \rightarrow p n)$ gives I=0,1 final two-nucleon state though 
neutron stimulated decay $(n \Lambda \rightarrow n n)$ gives only the I=1 
one.  The branching ratio of $\Gamma (n \Lambda \rightarrow n n) / \Gamma 
(p \Lambda \rightarrow p n)$ has been studied for several hypernuclei.  
Isospin structure studied by the ratio suggests dominance of I=1 
amplitudes over the I=0 ones, which contradicts calculations of meson 
exchange model where dominant tensor type interaction prefers I=0 final 
state.  Generally the NM-decay is assumed to be two body process due to a 
momentum transfer ($\sim$0.4 GeV/c) much larger than the Fermi momentum.  
However, experimental data are affected by the final state interaction and 
multi-nucleon mechanism due to the existence of other nucleons.  This 
situation obscures the assumption of the two-body process and makes 
comparison of measured branching ratios with theoretical models 
conceptually indirect.  

Protons from the proton stimulated decay are emitted asymmetrically 
with respect to the polarization of \lam\ hypernuclei.   Recently the 
asymmetry parameter has been studied by producing polarized \lam\ 
hypernuclei \cite{ajimura_nm,he5l_prop}.   The asymmetry parameter is due 
to interference of parity- conserving and violating amplitudes.  The 
relative phase of two amplitudes gives additional constraint on 
theoretical models for the process.  However, the precision of the 
experiment is limited by the final state interaction and magnitude of the 
polarization \cite{he5_pol,ogaito}.  

Recent sophisticated meson-exchange models of the weak BB interaction 
\cite{ramos,oset} have not completely solved an inherent problem for the 
NM-decay which is a difficulty to reproduce the transition rate and 
branching ratio simultaneously \cite{bhang}.  The meson exchange model 
is unable to account for the short range mechanism which is important in 
the NM-decay due to the large momentum transfer.   A quark exchange model 
is a natural one to incorporate the short range dynamics 
\cite{kisslinger,oka,inoue}, although in order to make realistic 
comparison with experiments the interplay between 
meson-exchange and quark-exchange mechanism has yet to be clarified.   
The NM-decay is the only tool to investigate the weak BB interaction 
beyond the NN interaction at present.  However, the initial \lam N state 
is constrained by the \lam\ hypernuclear structure and final two nucleon 
state is affected by final state interaction \cite{ramos_3n,oset}.   One 
wishes to derive two-body process of the NM-decay to understand the weak 
BB interaction though above facts make the derivation difficult.  

Recently it has been proposed that study of the NM-decay can be extended 
by the study of the inverse reaction (\pnpl ) \cite{tk_wein,hai,par,oka}.  
The Q value of the reaction is the mass difference between neutron and 
\lam\ (176 MeV) which requires 369 MeV proton kinetic energy for a free 
neutron target.  At this energy the strong interaction cannot produce 
strange particle; thus detection of \lam\ is the evidence of generation of 
strangeness by weak interaction.  The feasibility of the experiment is 
largely dependent on the cross section, for which several calculations 
have been carried out.  The observed NM-decay rate gives cross section of 
$\sim 10^{-39} \rm{cm}^2$\cite{tk_wein} at the corresponding kinematical 
region which is $\sim 10 \rm{MeV}$ above the threshold (E$_p 
\sim$400 MeV).  The theoretically calculated cross sections vary almost an 
order of magnitude $10^{-39} \sim 10^{-40} \rm{cm}^2$, depending on models 
used, reflecting our insufficient knowledge of the NM-decay 
\cite{hai,par,oka}.   The cross section is very small but the experiment 
is feasible with a sophisticated detector system under preparation
\cite{rcnp_prop,tk_qulen}.  

There are essential differences in the study of the \pnpl\ reaction 
although the reaction is the just inverse reaction of the NM-decay.  
In the inverse reaction one can employ a spin polarized proton beam and 
the polarization of \lam\ produced by the reaction can be measured by 
using the large asymmetry parameter $\alpha_- = 0.642 \pm 0.013$ 
\cite{pdg} of the $\Lambda \rightarrow p \pi^-$ decay.   The polarization 
of the proton beam can be either longitudinal or transverse with magnitude 
approaching unity.  This situation makes various spin observables 
measurable in the experiment.  
Such spin observables give interferences between amplitudes of 
parity-conserving and violating, spin-singlet and triplet, and 
isospin-singlet and triplet.  They will open new opportunity to study the 
weak YN interaction.  Here we derived the formulas of the spin 
observables and clarify the relation to the amplitudes commonly used in the 
study of the NM-decay.  

The general T-matrix of $p n \rightarrow p \Lambda$
can be expressed as follows assuming rotational invariance in the
center of mass system
\begin{eqnarray}
<s_{p'}s_{\Lambda};\bbox{p}'|\hat{T}|s_{p}s_{n};\bbox{p}>
& = & \sum_{S,S',L,L',J,L_z,L_z'}
      (1/2 s_{p}' 1/2 s_{\Lambda}|S' S_z')
      (L' L_z' S' S_z'|J M) Y_{L',L_z'}(\hat{p}') \nonumber \\
& &  \times     (1/2 s_{p} 1/2 s_{n}|S S_z)
      (L L_z S S_z|J M) Y_{L,L_z}^*(\hat{p}) \nonumber \\
& &  \times  4\pi<(L'S')JM;p'|T|(LS)JM;p>
\end{eqnarray}
Here $s_{p}',s_{\Lambda}, s_{p}$ and $s_{n}$ are baryon spins.  
$\bbox{p},\bbox{p}'$ are the momenta of initial and final proton.
In order to calculate the polarization observables, we introduce the 
following density matrix $\rho$

\begin{eqnarray}
\rho_{i} = \frac{1 + \bbox{\sigma}_{i}\cdot \bbox{P}_i}{2}
\end{eqnarray}
where $i$ stands for proton($p$) or $\Lambda$($\Lambda$) and $\bbox{P}_i$
is polarization vector of particle $i$.
The differential cross section can be simply calculated by taking the
following trace of the baryon spins.
\begin{eqnarray}
\frac{d\sigma}{d\Omega} \sim
 Tr [ \rho_{\Lambda} \hat{T} \rho_{p} \hat{T}^{\dagger} ]
\end{eqnarray}

We restrict our treatment to s-wave production for the $p \Lambda$ 
states to focus our discussion on the relation to the NM-decay.   
Accordingly the maximum angular momentum is $J=1$.  By this truncation one 
can have a transparent representation of the observables by the following 
well known 6 amplitudes.
\begin{eqnarray}
a & = & <^1S_0|\hat{T}|^1S_0,I=1,P=+>  \nonumber \\
b & = & <^1S_0|\hat{T}|^3P_0,I=1,P=->  \nonumber \\
c & = & <^3S_1|\hat{T}|^3S_1,I=0,P=+>  \nonumber \\
d & = & <^3S_1|\hat{T}|^3D_1,I=0,P=+>  \nonumber \\
e & = & <^3S_1|\hat{T}|^1P_1,I=0,P=->  \nonumber \\
f & = & <^3S_1|\hat{T}|^3P_1,I=1,P=->   \label{tmat1}
\end{eqnarray}
where isospin($I$) and parity($P$)
 of the initial pn system are explicitly written.  
Using the above amplitudes, the spin structure of the T-matrix is given 
as follows  similar to the one for the NM-decay by Block and 
Dalitz \cite{block} as 
\begin{eqnarray}
\hat{T} & =&  a \frac{1-\bbox{\sigma}_p\cdot\bbox{\sigma}_{\Lambda}}{4}
  - b\frac{1-\bbox{\sigma}_p\cdot\bbox{\sigma}_{\Lambda}}{8}
(\bbox{\sigma}_p-\bbox{\sigma}_{\Lambda})\cdot\hat{\bbox{p}}
  +   c \frac{3+\bbox{\sigma}_p\cdot\bbox{\sigma}_{\Lambda}}{4} \nonumber \\
& &   + d\frac{1}{2\sqrt{2}}
     (3\bbox{\sigma}_p\cdot\hat{\bbox{p}}
       \bbox{\sigma}_{\Lambda}\cdot\hat{\bbox{p}}-
       \bbox{\sigma}_p\cdot\bbox{\sigma}_{\Lambda})
  +e \frac{\sqrt{3}(3+\bbox{\sigma}_p\cdot\bbox{\sigma}_{\Lambda})}{8}
      (\bbox{\sigma}_p-\bbox{\sigma}_{\Lambda})\cdot\hat{\bbox{p}}
  - f\frac{\sqrt{6}}{4} (\bbox{\sigma}_p+\bbox{\sigma}_{\Lambda})
               \cdot\hat{\bbox{p}}
\end{eqnarray}

The differential cross section is written as follows,
\begin{eqnarray} 
\frac{d\sigma}{d\Omega} & = &
  \frac{d\sigma}{d\Omega}|_{unpol.}[
 1 + \bbox{P}_{p}\cdot \hat{\bbox{p}}\,  A_p
   + \bbox{P}_{\Lambda}\cdot \hat{\bbox{p}}\,  A_{\Lambda}  \nonumber \\
& &
 +
   \bbox{P}_{p}\cdot\hat{\bbox{p}}
   \bbox{P}_{\Lambda}\cdot\hat{\bbox{p}}\, A_{p\Lambda}^{L}
 +
\bbox{P}_{p}\times\hat{\bbox{p}}
         \cdot
\bbox{P}_{\Lambda}\times\hat{\bbox{p}}\,  A_{p\Lambda}^{T}
 +
\bbox{P}_{p}\times\bbox{P}_{\Lambda}\cdot\hat{\bbox{p}}\,  
A_{p\Lambda}^{T'}]
\end{eqnarray}

Coefficients of each term are represented as follows, 

\begin{eqnarray}
  A & = & |a|^2 + |b|^2 + 3 [ |c|^2 + |d|^2 + |e|^2 + |f|^2]  \\
  A_{p} & = &
   2 \sqrt{3} Re [ - a b^*/\sqrt{3} + e ( c - \sqrt{2}d)^*
                  -  f (\sqrt{2}c + d)^* ]/A \\
  A_{\Lambda} & = &
   2 \sqrt{3} Re [ - a e^* + b(c -\sqrt{2}d)^*/\sqrt{3}
                           - f(\sqrt{2}c + d)^*]/A  \\
  A_{p\Lambda}^{T} & = &
   Re [ - \sqrt{2}a (\sqrt{2}c +d)^*
        +2|c|^2 - \sqrt{2} cd^* - 2|d|^2 
        +  \sqrt{6}f(b+\sqrt{3}e)^* ]/A \\
  A_{p\Lambda}^{L} & = &
     Re [ - 2a (c - \sqrt{2}d)^*
         + 2|c|^2 + 2\sqrt{2}cd^* + |d|^2 +  2\sqrt{3}be^* + 3|f|^2 ]/A \\
  A_{p\Lambda}^{T'} & = &
    \sqrt{6} Im [ a f^* - (b/\sqrt{3} + e)(\sqrt{2}c + d)^*
                         + f(c - \sqrt{2} d)^*]/A 
\end{eqnarray}

In the present reaction we have 6 observables including cross section 
($A$).  It 
has to be stressed that all 6 observables are measurable in the 
experiment.  

$A_p$ and $A_\Lambda$ are correlations that violate parity.  
Experimentally $A_p$ is obtained by measuring differences of the cross 
section for longitudinally polarized beams.  
\begin{equation} 
A_p = {\sigma (h_p=1)- \sigma (h_p=-1)\over \sigma (h_p=1)+ \sigma (h_p=-1)}
\label{eq:ana_p} 
\end{equation} 
where helicity is defined as $h_p= \bbox{\sigma}_p\cdot \hat{\bbox{p}}$ 
and $h_p \sim 1$ is experimentally achievable.  Similarly, the polarization 
of \lam\ in the beam direction gives $A_\Lambda$.  

There are 9 interference terms ($3(P=+) \times 3(P=-)$) that violate 
parity.  $A_p$ has no interference term between $J=0$ and $J=1$ states 
because spin average is taken in the final $p \Lambda$ system.  
One can thus see 5 terms (1$(J=0) + 2 \times 2(J=1)$) in the equation.  

$A_{\Lambda}$ gives the polarization of \lam\ in the proton beam direction 
that violates parity.  It cannot be described by the definite isospin of 
the two nucleons in the initial state because an exchange of proton and 
neutron is equivalent to the parity when we average spin of the initial 
$pn$ system.  Thus $A_{\Lambda}$ gives the interference between $I=0$ and 
$I=1$ matrix elements and 4 interference terms ($2 (I=0) + 2 (I=1)$) 
between the same isospin disappear.  

The asymmetry parameter in the NM-decay ($\alpha_p$) is the only 
interference term that has been experimentally measured so far 
\cite{tk_kek,ajimura_nm,he5l_prop}.  Spin polarized hypernuclei has 
asymmetric emission of protons from the proton stimulated NM-decay 
represented as 
\begin{eqnarray} 
W(\theta ) = 1 + \alpha_p cos \theta . 
\end{eqnarray}
$\alpha_p$ has been given as $2 \sqrt{3}f (\sqrt{2} c + d)/A$ assuming 
the initial \lam N system is in a relative s-wave \cite{bando_mod}.  It is 
essentially equivalent to 
$A_{\Lambda}$ except for difference in the initial- and final-state 
interactions.  Here we skip a subtle issue related to the convention of 
phase which is irrelevant to the present argument.  It is noticed that 
$A_{\Lambda}$ includes the contribution of the singlet initial 
state, which is missing in the formula of $\alpha_p$.  
It is obvious that singlet state alone gives no asymmetry.  However, the 
\lam N system buried in a hypernucleus can be singlet- and triplet-states 
whose interference terms make the formula of $\alpha_p$ equivalent to 
that of $A_{\Lambda}$.   

There are three double polarization observables.  Spin polarization 
is classified into transverse- and longitudinal-types.  
$A_{p\Lambda}^{T}$ is the correlation of transverse polarization.  
The $|c|^2$ term ($^3S_1 \rightarrow ^3S_1$) keeps initial polarization 
although the $|d|^2$ term ($^3D_1 \rightarrow ^3S_1$) flips it.  No parity 
violation appear in this correlation thus 
interference terms are restricted to the same parity.  

$A_{p\Lambda}^{T'}$ is a parity and T (time reversal) violating 
observable.   It corresponds to generation of the \lam\ polarization 
in the direction defined by transverse proton polarization and 
proton momentum.  
The T violating correlation being searched for in the 
$K^+ \rightarrow \pi^0 \mu^+ \nu_\mu$ decay is $\bbox{p}_\pi \times \bbox{p}_\mu 
\cdot \bbox{P}_\mu$ (P even and T odd) \cite{e246}.  No search has been 
carried out for flavor changing baryon baryon interaction.  So far it is 
known that theories that violate T invariance also violates 
parity\cite{gudokov}, making $A_{p\Lambda}^{T'}$ 
(P odd and T odd) correlation interesting.  These consideration makes the 
search valuable even though expected precision is inferior to the kaon 
decay\cite{tk_wein,tk_qulen}.  It is known that spurious T violation is 
seen in the $\Lambda \rightarrow p \pi^-$ decay due to final state 
interaction.  The final state interaction has to be evaluated for the 
T violation experiment.  This is left for the future study.  

$A_{p\Lambda}^{L}$ is the correlation for the longitudinal polarization.  
It corresponds to the spin-flip probability in the beam direction and is 
a parity-conserving correlation.  However, the correlation would not give 
a deep insight to the \pnpl\ reaction that has the large parity violation.   
Experimental data can be transparently related to the relevant amplitudes in 
the helicity representation.  We have 6 independent amplitudes
$T(h_{p'},h_{\Lambda};h_p,h_n)$, which are given in terms 
of multipole amplitudes as
\begin{eqnarray}
 T( 1, 1; 1, 1) & = \frac{c_+ - \sqrt{3}f}{\sqrt{2}} \\ 
 T(-1,-1;-1,-1) & = \frac{c_+ + \sqrt{3}f}{\sqrt{2}} \\ 
 T(-1, 1; 1,-1) & = \frac{-a + c_- + b_+}{2} \\
 T( 1,-1;-1, 1) & = \frac{-a + c_- - b_+}{2} \\ 
 T( 1,-1; 1,-1) & = \frac{ a + c_- - b_-}{2} \\ 
 T(-1, 1;-1, 1) & = \frac{ a + c_- + b_-}{2} , 
\end{eqnarray}
where
\begin{eqnarray}
c_+ & = & \sqrt{2}c + d \\
c_- & = &  c - \sqrt{2}d \\
b_+ & = &  b + \sqrt{3}e \\
b_- & = &  b - \sqrt{3}e.
\end{eqnarray}
Here all $h_i$ represent spins of baryons in the direction of incoming proton 
momentum.  
Using longitudinal polarization of proton and \lam , we can determine 4
independent combinations of the absolute magnitude of amplitudes,
which can also be represented by combinations of $A, A_p, A_{\Lambda}$ and
$A_{p\Lambda}^L$.  One can conveniently obtain expressions of observables 
using longitudinal polarization by interference terms. 
For example polarization of $p$ and $\Lambda$, which are anti-parallel gives
\begin{eqnarray}
\frac{\sigma(h_p=1,h_{\Lambda}=-1)- \sigma(h_p=-1,h_{\Lambda}=1)}
     {\sigma(h_p=1,h_{\Lambda}=-1)+ \sigma(h_p=-1,h_{\Lambda}=1)}
 & = &
\frac{|T(1,-1; 1,-1)|^2 - |T(-1, 1;-1,1)|^2}
{|T(1,-1; 1,-1)|^2 + |T(-1, 1;-1,1)|^2} \nonumber \\
 & \sim  & Re[(a+c_-)b_-^*] \label{cros1}
\end{eqnarray}
It is interesting that we can remove $f$ in Eq. (\ref{cros1}),
 which was suggested to be dominant from the phenomenological 
analysis \cite{block}, though it 
may not be so large in the meson-exchange models.  
Here we have not discussed terms relevant to polarization of target 
neutron since feasibility of the experiment is currently questionable.  

Here we restricted our discussion to the relative s-wave which corresponds 
to proton kinetic energy of $\sim$400 MeV, although it can be extended to 
include higher partial waves.  The present kinematic regime is selected 
because the NM-decay rate can give an order of magnitude estimation of the 
cross section and the obtained result should be useful to understand the 
NM-decay in detail.  However, one can study energy dependence of the 
\pnpl\ reaction safely up to proton energy around 680 MeV where \lam\ and 
kaon pair production become possible only for the limit of infinitely 
heavy hypernuclei.   The study of energy dependence will give information 
on the general structure of the weak YN interaction.   The study of the 
p-wave component of the reaction is particularly interesting.  If the 
meson exchange models is insufficient to describe the NM-decay only in the 
short range region, the models should well describe the long range part 
which is naturally associated with the p-wave part of the interaction.  

In summary, we derived formulas of spin observables in the \pnpl\ 
reaction.   Those observables are useful not only to study the NM-decay 
of \lam\ hypernuclei but also to study the weak BB interaction generally.  
The spin polarization of the incident proton beam can be large and 
precisely given.  The polarization of \lam\ is also well determined 
experimentally.  The spin observables are affected little by nuclear 
effects which limit study of the NM-decay.  The spin and isospin of the 
\lam N system is determined by the hypernuclear wave functions for the 
NM-decay though 
the \pnpl\ reaction has no such limitation.  The reaction is shown to 
be useful for the study of the weak BB interaction.  

T.K is grateful to professors M. Oka, A. Gal, A. Ramos, C. Bennhold and 
E. Oset for discussions on this subject.  T.K. is grateful to members of 
the experimental collaboration with whom this experiment is under 
preparation.  The authors are grateful to Dr. R. E. Chrien for careful 
reading of this manuscript.  This work has been partly supported by the 
Grant-in-Aid for Scientific Research in Priority Areas (Strangeness Nuclear 
Physics) for the Ministry of Education, Science, Sports and Culture of 
Japan.

\bibliographystyle{unsrt}

\end{document}